\documentclass[12pt]{article}

\usepackage{amssymb}
\usepackage[T1]{fontenc}
\usepackage{amssymb,amsthm,amsmath,mathrsfs,bm}

\begin{document}

\title {N=2 Supercomplexification of the Korteweg-de Vries, Sawada-Kotera and Kaup-Kupershmidt Equations}

\author{Ziemowit Popowicz}

\maketitle

\begin{center}{
Institute of Theoretical Physics, University of Wroc\l aw,
Wroc\l aw pl. M. Borna 9, 50-205 Wroc\l aw Poland, ziemek@ift.uni.wroc.pl}
\end{center}
\vspace{0.8cm}

\begin{abstract}
The supercomplexification  is a special method of $N=2$ supersymmetrization of the integrable equations in which the bosonic sector can be reduced to the 
complex version of these equations. The $N=2$ supercomplex Korteweg-de Vries, Sawada-Kotera and Kaup-Kupershmidt equations are defined and investigated. 
The common  attribute of the supercomplex equations is appearance of the odd Hamiltonian structures and superfermionic conservation laws. 
The odd bi-Hamiltonian structure, Lax representation and superfermionic conserved currents  for new $N=2$  supersymmetric Korteweg-de Vries  equation 
and for Sawada-Kotera one, are given. 
\end{abstract}


\section{Introduction}

Integrable Hamiltonian systems occupy an important place in diverse branches of theoretical physics as exactly solvable models of fundamental 
physical phenomena ranging from nonlinear hydrodynamics to string theory \cite{Fad, dic}.  There are many approaches to investigate these systems as for example 
the Lax approach, the construction of the recursion operator and bi-Hamiltonian structure or checking  the B\"{a}cklund and Darboux transformations \cite{Blask}. 

On the other hand, applications of the supersymmetry (SUSY) to the soliton theory
provide  a possibility of  generalization of  the integrable systems. The supersymmetric integrable equations [4-19] have drawn a lot of attention  for a variety of
reasons. In order to create a supersymmetric theory, we have to add to a system of $k$ bosonic
equations $kN$ fermionic and $k(N-1)$  bosonic fields ($k=1,2 \dots N=1,2 \dots$) in such a way that the
final theory becomes SUSY invariant. A bonus of this method is, that the so called bosonic sector of the supersymmetrical equations  when $N \geq 2$,  leads us to 
a new system of interacting fields. For example,  the Virasoro algebra \cite{Gerva1} and some of its extensions can be related to the second Hamiltonian structure of the 
Korteweg-de Vries (KdV) and KdV-like equations. This Hamiltonian structure is given by the set of Poisson brackets for the fundamental fields representing the 
Virasoro algebra. Now, starting from the supersymmetric generalization of the Virasoro algebra and the corresponding Hamiltonian structure,  the  $N=1,2$ supersymmetric 
extensions of the classical equations have been obtained \cite{Mat,LMat,Lmm}.

There are many methods of supersymmetrization of integrable system as, for example, to start simply from the supersymmetric version of the Lax operator or consider 
the supersymmetric version of the Hamiltonian structure. 
Interestingly,  during the process of supersymmetrization many unexpected, but typical supersymmetric  effects occurred. 
In particular,  the roots for the SUSY Lax operator are not uniquely defined \cite{Opop}, 
the non-local conservation laws \cite{Das}   and the odd Hamiltonian structure appear \cite{Pop2,Pop4}.

The idea of introducing  odd Hamiltonian structure is not new. Leites noticed \cite{Leits}  that in the superspace,  one can 
consider both the even and odd sympletic structures, with even and odd Poisson brackets respectively. The odd brackets,  also known
as antibrackets,  have drawn some interest in the context of BRST formalism in the Lagrangian framework \cite{Bat}, 
in the supersymmetrical quantum mechanics \cite{Volk}, and in the classical mechanics \cite{Kup2}. 

Becker  and Becker  in \cite{Backer} proposed  the supersymmetric KdV equation in the form 
\begin{equation} \label{bak}
 \Phi_t = \Phi_{xxx} + 6({\cal D}\Phi) \Phi_x, 
\end{equation}
where $\Phi$ is a superfermionic $N=1$  function. If we replace $u$ in the KdV equation by $({\cal D}\Phi)$  then we obtain the equation (\ref{bak}).  
However, as a result,  the bi-Hamiltonian structure becomes the odd one.

The  $N=2$ supersymmetric generalization of KdV equation  
\begin{equation}
\Phi_t = \Phi_{xxx} + 6({\cal D}_1{\cal D}_2 \Phi) \Phi_x,
\end{equation} 
was considered in \cite{Pop2}. Similarly to the Becker and Becker  equation (\ref{bak}) this equation 
possess  the odd bi-Hamiltonian structure.
However the Lax representation for this equation has been not presented. 

In this paper, we generalize the $N=1$ substitution $u=>({\cal D}\Phi) $ to the $N=2$ supersymmetric case,  assuming that 
\begin{equation} \label{sup}
 u => (k_1 ({\cal D}_1{\cal D}_2 \Phi) + k_2\Phi_x) + i(k_3 ({\cal D}_1{\cal D}_2 \Phi) + k_4\Phi_x)),
\end{equation}
where $i^2 =-1$ and $k_j,  j=1,...,4$ are arbitrary constants. We investigate the Hamiltonian structure, Lax representation and conservation laws for the obtained 
equations after such substitution. We call such substitution $BN=2$ supercomplexification. An unexpected feature of this supercomplexification is that if 
we directly substitute the ansatz eq.(\ref{sup}) to the conserved currents of the KdV equation, then the currents are no longer  conserved currents, 
in  contrast to the $N=1$ case. As we show,  such a supercomplexification leads us to the odd Hamiltonian structures and to the superfermionic conserved currents.

We investigate  supercomplexifications  of three equations: the KdV, Sawada-Kotera (S-K) and Kaup-Kupershmidt (K-K) equations. 
For all these equations,  we fix the arbitrary constants in such a way that their bosonic sector could be transformed to the complex version of the KdV, S-K, K-K 
equations. This procedure justifies the name of supercomplexification. 
The odd bi-Hamiltonian structure, Lax representation and superfermionic conserved currents  for new $BN=2$  supersymmetric Korteweg-de Vries  equation are given.
For the $BN=2$ supercomplex Sawada-Kotera equation the Lax representation, odd bi-Hamiltonian structure and superfermionic conserved currents are defined. 
The $BN=2$ supercomplex Kaup-Kupershmidt equation is defined,  for which the odd bi-Hamiltonian structure is presented  with its superfermionic conserved currents. 

All calculations used in the paper have been carried out  with  the help of computer program Susy2 \cite{sus}.

The paper is organized as follows. In the first section,  the notation used in the non-extended and in extended supersymmetry is explained.  Section  2 contains 
description of the non-extended $N=1,BN=1$ and extended $N=2,BN=2$ supersymmetric KdV equation. Section  3 and section 4   treat the 
non-extended $N=1,BN=1$ and extended $N=2,BN=2$ supersymmetric Sawada-Kotera and Kaup-Kupershmidt equations. The last section  is the conclusion. 

\section{Notation used in the supersymmetry}

In the non-extended $N=1$ supersymmetric theory, we deal with the odd and even variables. These variables  are 
joined in the multiplet as $
 \Phi = \xi + \theta u $ or as $ \Upsilon = w + \theta \zeta$
where   $\xi=\xi(x,t), \xi^2 =0 , \zeta =\zeta(x,t),\zeta^2 =0 $ are odd functions while   $u=u(x,t), w=w(x,t)$ are  even functions,
and $\theta$ is  Majorana spinor such that $\theta^2 =0$. In other words  $\theta$ is the odd coordinate.   
The $\Phi$ is called  superfermionic function while $\Upsilon$ a superbosonic one. 

The supersymmetric derivative ${\cal D}$ is defined as 
\begin{equation} 
 {\cal D} = \frac{\partial }{\partial \theta} + \theta \partial, \hspace{2cm} {\cal D}^2 =\partial.  
\end{equation}
The  symbolic integration over the odd variables is defined as 
\begin{equation} 
 \int ~ d\theta =0, ~~~~~~ \int ~~ d\theta ~ \theta =1. 
\end{equation}

In the extended supersymmetry $N=2$ case we deal with more complicated superfermionic or superbosonic functions which are 
defined by
\begin{equation} 
 \Phi = w +  \theta_1 \xi_1 + \theta_2 \xi_2 + \theta_1 \theta_2 u, ~~~~~
 \varUpsilon = \zeta_1  + \theta_1 h + \theta_2 k + \theta_1 \theta_2 \zeta_2,
\end{equation}
where $w,u,k,h$ are even functions,  $ \xi_1, \xi_2, \zeta_1,\zeta_2, \xi_i^2=0, \zeta_i^2=0, \xi_2 \xi_1 = -\xi_1 \xi_2 , \zeta_2\zeta_1=-\zeta_1\zeta_2 $ 
are odd functions which take values in the Grassman algebra,  
$\theta_1$ and $\theta_2$ are two different Majorana spinors, odd coordinates,  such that $ \theta_i^2=0, \theta_2 \theta_1 =- \theta_1\theta_2$. 
$\Phi$ is the superboson function  while  $\varUpsilon$ is   superfermionic function. 

The  supersymmetric derivatives and symbolic integrations are defined as 
\begin{eqnarray} \label{sprop}
&& {\cal D}_i = \frac{\partial }{\partial \theta_i} + \theta_i \partial , ~~~, {\cal D}_i^{-1} = {\cal D}_i\partial^{-1}, ~~~i=1,2,  \\  
&& {\cal D}_1^2 = {\cal D}_2^2 =\partial, ~~~ {\cal D}_1 {\cal D}_2 + {\cal D}_2 {\cal D}_1 =0,  \\ 
 && \int \theta_2 \theta_1 d \theta_1 d \theta_2 =1, ~~~~~~ \int d \theta_1 d \theta_2 =0,
\end{eqnarray}
where negative  power $\partial^{-1}$ of formal integration is defined as 
\begin{equation} 
  \partial^{-1} a = a\partial^{-1} - a_x\partial^{-2} + a_{xx}\partial^{-3} \mp \ldots, ~~~
  a\partial^{-1} = \partial^{-1}a + \partial^{-2}a_x +\partial^{-3}a_{xx} + \ldots.
\end{equation}
From the formulas (\ref{sprop}) follows  $\int d \theta_1 d \theta_2 dx ({\cal D}_i \varGamma) = 0, i=1,2$
for an arbitrary superfunction $\varGamma$ which  vanishes in $\pm  \infty$.

For the supersymmetric extensions of the models discussed in what follows, the Lax operators may be regarded as the element of 
the algebra of super-differential operators $\cal G$.
For $N=1$ we have
\begin{equation} 
{\cal G} :=\{ \sum_{k=-\infty}^{\infty} (a_k + \Phi_k {\cal D})\partial^{k} \},
\end{equation}
and for $N=2$ 
\begin{equation} 
{\cal G} :=\{ \sum_{k=- \infty}^{\infty} (b_k + \beta _k {\cal D}_1 + \gamma_k{\cal D}_2 + a_k {\cal D}_1{\cal D}_2) \partial^{k} \}.
\end{equation}

The supersymmetric algebra ${\cal G}$ possess three invariant subspaces defined by the following projections $P$  onto the subspaces of  ${\cal G}$. 
\begin{equation} 
P_{\geq 0} ({\cal G}) = {\cal G}_{\geq 0} =  \{ \sum_{k=0}^{\infty} (f_k + \beta_k{\cal D}_1 + \gamma_k{\cal D}_2 + g_k{\cal D}_1{\cal D}_2)\partial^k  \},
\end{equation}

\begin{equation} 
P_{\geq 1} ({\cal G}) = {\cal G}_{\geq 1} =  \{ \sum_{k=1}^{\infty} (f_k + \beta_k{\cal D}_1 + \gamma_k{\cal D}_2 + g_k{\cal D}_1{\cal D}_2)\partial^k + 
	\alpha{\cal D}_1 + \delta{\cal D}_2 + h{\cal D}_1{\cal D}_2 \},
\end{equation}

\begin{eqnarray} \label{trace} 
P_{\geq 2} ({\cal G})&=& {\cal G}_{\geq 2} =  \{ \sum_{k=2}^{\infty} (f_k + \beta_k{\cal D}_1 + \gamma_k{\cal D}_2 + g_k{\cal D}_1{\cal D}_2)\partial^k + \\ \nonumber 
&& \hspace{2cm} 	(f_1{\cal D}_1 + f_2{\cal D}_2 + f_3{\cal D}_1{\cal D}_2)\partial + \delta {\cal D}_1 {\cal D}_2  \}. 
\end{eqnarray}

The subscript $L_{\geq 0}, L_{\geq 1}$ in what follows  denotes the projection $P_{\geq 0}(L), P_{\geq 1}(L)$ . 

The supersymmetric algebra ${\cal G}$ is endowed with non-degenerate ``trace form'' given by the residues
\begin{eqnarray} \label{trace}
 tr(L) &=& Res( \sum_{k < \infty} (a_k + \Phi_k {\cal D}) \partial^k) = \int ~dx d \theta \Phi_{-1}, \\ 
 tr(L) &=& Res( \sum_{k < \infty} (b_k + \beta_k {\cal D}_1 + \gamma_k {\cal D}_2 + a_k {\cal D}_1{\cal D}_2) \partial^k) = \int ~dx d \theta_1 d \theta_2  a_{-1},
\end{eqnarray}
for $N=1$ and $N=2$ respectively. 
This ``trace form '' is used in the theory of integrable systems because from the knowledge of the Lax operator it is possible to obtain the conserved currents. 

Similarly to the classical case it is possible to construct the generalized supersymmetric Lax representation $ \frac{d }{d t_q} L = 
\big [ P_{\geq k}(L^q),L \big ]$ where $k=0,1,2$. Restriction to  $k=0$  yields to the supersymmetric Gelfand-Dikii hierarchy of equations. Restriction to 
$k=1$ yields the supersymmetric  nonstandard Lax representation while for $k=2$ to the supersymmetric Harry Dym hierarchy. 

As usual in the case of extended supersymmetry $N > 1$,  we assume the invariance of the considered model  under  change the odd variables. 
It means that we always assume the invariance under the replacement of the supersymmetric derivatives 
${\cal D}_1 \rightarrow  -{\cal D}_2, {\cal D}_2 \rightarrow {\cal D}_1$ and denote this transformation  as $O_2$. For example we have $O_2({\cal D}_1\Phi) = 
-({\cal D}_2 \Phi)$.

\section{N=1,BN=1,N=2 and BN=2   susy  KdV }

The  Korteweg-de Vries equation is defined as 
\begin{eqnarray} \label{kkdv} 
&& u_t =  u_{xxx} + 6 uu_x =  J \frac{\delta  H_1}{\delta u} = P  \frac{\delta H_2}{\delta u}, \\ 
&&  J=\partial, ~~~~~~ P =  \partial^{3} + 2\partial u +2 u \partial,  \\ \nonumber
&& H_1=\frac{1}{2} \int dx~(uu_{xx} +4 u^3), ~~~~H_2 =\frac{1}{2} \int dx~ u^2,
\end{eqnarray}
and is obtained from the  Lax representation 

\begin{eqnarray} \label{laxik} 
 L &=& \partial^2 + u, \hspace{1cm} L_t=[ L,L^{3/2}_{\geq 0} ].  \\ \nonumber 
\end{eqnarray}

The subscript $\geq 0$ in  $ (L^{3/2})_{\geq 0}$ denotes the purely differential  part of $L^{3/2}$.
The KdV equation  is integrable, has bi-Hamiltonian structure and possesses infinite number of bosonic conserved currents. 
The Lax operator Eq.(\ref{laxik}) generates the conserved currents as
\begin{equation} \label{slad} 
 H_n=tr(L^{(2n+1)/2}) = tr(\partial^{(2n+1)/2} + \sum_{i=-\infty}^{(2n-1)/2} a_i\partial^{i}) = \int dx a_{-1} = \int dx h_n.
\end{equation}

Let us recall the relationship between the Poisson bracket 
\begin{equation} 
\{ u(x),u(y) \} = \frac{1}{2} [ \partial^{3} + 2\partial u +2 u \partial ] \delta(x-y)
\end{equation} 
and the Virasoro algebra \cite{Gerva,Gerva1}. Fourier expansion of 
\begin{equation} 
 u(x) = - \frac{12}{c}\sum_{n=-\infty}^{\infty} L_n e^{-inx} + \frac{1}{2}
\end{equation}
where $c$ is a constant leads us to the classical form of the Virasoro algebra
\begin{equation} 
 \{ L_n,L_m \} = (n-m)L_{n+m} +  \frac{c}{12}\delta(n+m,0).
\end{equation}

\subsection{N=1, BN=1 supersymmetric KdV equation}

The  $N=1$ supersymmetric KdV equation is  obtained from the 
Lax representation 
\begin{eqnarray} 
&&  (L=\partial^2 + {\cal D}\Phi)_t=[ L, L^{3/2}_{>0}] ~~ => ~~
 \Phi_t =  (\Phi_{xx} + 3({\cal D}\Phi) \Phi)_x ,  \\ 
&& \xi_t = (\xi_{xx} + 3\xi u)_x , ~~~~ u_t = (u_{xx} + 3u^2 + 3\xi_x\xi)_x. 
\end{eqnarray}
This equation has been thoroughly investigated in many papers \cite{ Krasnal, Kup1, Manin}. 

The $BN=1$ supersymmetric KdV equation is generated by the  Lax representation 
\begin{eqnarray} \label{becker} 
 && (L= \partial^2 + ({\cal D}\Phi))_t=[L,L^{3/2}_{\geq 0}]  ~~~ => ~~~
 \Phi_t =  \Phi_{xxx} + 6({\cal D}\Phi) \Phi_x,  \\ 
&& \xi_t = \xi_{xxx} + 6\xi_x u, ~~~~  u_t = (u_{xx} + 3u^2)_x. 
\end{eqnarray}
It has a triangular form, $u_t$ does not contain the odd function,  but it is a very interesting equation from   integrability 
and supersymmetry point if view. This equation has been first considered by Becker and Becker \cite{Backer}  and it was named 
later as B extension of supersymmetric KdV equation
 \cite{Das}.

This system possesses  infinite number of the superfermionic conservation laws. For example,  
\begin{eqnarray} 
 H_{3.5} &=& \frac{1}{2} \int dx ~ d\theta ~\Phi \Phi_x = -\int dx~\xi u_x, \\ 
 H_{5.5} &=& \frac{1}{2} \int dx ~ d\theta ~\Phi ~(\Phi_{xxx} + 4\Phi_x ({\cal D}\Phi)) =- \int dx ~\xi(u_{xxx} + 6u_x u). 
\end{eqnarray}
where lower index in $H$ denotes the dimension of the expression. Assuming that $deg(A)$ denotes the dimension of $A$, we have 
\begin{eqnarray} 
 && deg~ u=2, ~~~deg~ \Phi=3/2, ~~~
deg ~{\cal D}=\frac{1}{2},~~~~~ deg~ \theta =-\frac{1}{2}, \\ \nonumber 
&& deg ~\xi = 3/2, ~~~ deg~ x =-1,~~~ deg ~\partial =1.
\end{eqnarray} 

The variational derivative $\frac{\delta }{\delta \Phi} $  of these superbosonic conservation laws will be a superfermionic function. Therefore,  our 
bi-Hamiltonian structure is living in the superfermionic space and hence the Hamiltonians operators  should be symmetrical operators. 

The bi-Hamiltonian structure is easy to obtain using the formula (\ref{kkdv}) in which we assume $u =({\cal D}\Phi) ~~ => ~~ \Phi = ({\cal D}^{-1}u) $ from which follows 
\begin{eqnarray} \label{tryk} 
&& \Phi_t = {\cal P}  \frac{\delta H_2}{\delta u} = {\cal D}^{-1} ( \partial^3 + 2\partial ({\cal D}\Phi) + 2  ({\cal D}\Phi) \partial ) 
 {\cal D}^{-1} \frac{\delta H_{3.5}}{\delta  \Phi} = \\ 
 && (\partial^2 +  4({\cal D}\Phi) + 2\partial^{-1} {\cal D} \Phi_x + 2\Phi_x \partial^{-1} {\cal D}) \frac{\delta H_{3.5}}{\delta \Phi} 
  = {\cal D}^{-1} \partial {\cal D}^{-1} \frac{\delta  H_{5.5}}{\delta \Phi}   = \frac{\delta H_{5.5}}{\delta \Phi}. 
\end{eqnarray}
The operator  ${\cal  P} $ is connected with the Poisson bracket
\begin{equation} 
 \{ \Phi(x,\theta),\Phi(y,\theta^{'})\} = {\cal P}  \delta(x-y)(\theta - \theta^{'}),
 \end{equation}
and is  rewritten in the components as 
\begin{eqnarray} 
 && \{u(x), u(y) \} = 4\xi_x \delta(x-y), \\ \nonumber 
 && \{u(x),\xi(y) \} = (\partial^2 + 2\partial^{-1}u_x + 4u) \delta(x-y), \\ \nonumber
 && \{ \xi(x),\xi(y) \} = 2\xi_x \delta(x-y). 
\end{eqnarray}

\subsection{N=2 and  BN=2 supersymmetric   KdV Equation}

We have three different $N=2$ supersymmetrical  extensions of the KdV equation 
\begin{equation} \label{trzykdv}
 \Phi_t = \big ( - \Phi_{xx} + 3\Phi ({\cal D}_1{\cal D}_2 \Phi) + \frac{1}{2} (\alpha -1)({\cal D}_1 {\cal D}_2\Phi^2) + \alpha \Phi^3 \big )_x,   
\end{equation}
where $\alpha = 1,-2,4$. 

All these equations possess the bi-Hamiltonian structure \cite{Opop,Sorin}. For example the second Hamiltonian structure for the equation (\ref{trzykdv}) is 
\begin{eqnarray} 
 \Phi_t &=& \Lambda \frac{\delta }{\delta \Phi}
 \frac{1}{2} \int ~dx d\theta_1 d\theta_2 (\Phi {\cal D}_1{\cal D}_2\Phi + \frac{\alpha}{3} \Phi^3), \\ \nonumber 
\Lambda &=&  {\cal D}_1{\cal D}_2\partial + 2 \partial \Phi + 2 \Phi \partial - {\cal D}_1 \Phi {\cal D}_1 - 
 {\cal D}_2 \Phi {\cal D}_2,
\end{eqnarray} 
and is connected with the Poisson bracket
\begin{equation} \label{pois1}
 \{ \Phi(x,\theta_1, \theta_2),\Phi(y,\theta_1^{'}, \theta_2^{'})\} = \Lambda \delta(x-y)(\theta_1 - \theta_1^{'})(\theta_2 - \theta_2^{'}).
\end{equation}
This formula could be rewritten in the components 
\begin{eqnarray} \label{comvir}
 && \{u(x), u(y) \} = (-\partial^3 + 4u\partial + 2u_x)\delta(x-y), \\ \nonumber 
 && \{u(x),\xi_i(y) \} = (3\xi_i + \xi_{i,x})\delta(x-y), \\ \nonumber 
 && \{ u(x),w(y) \} =  2w \delta(x-y)_x, \\ \nonumber 
 && \{ \xi_i(x),\xi_j(y) \} = [ \delta_{i,j}(\partial^2 - u) + \epsilon_{i,j}(2w\partial + w_x) ] \delta(x-y), \\ \nonumber 
 && \{ w(x),w(y) \} = \delta(x-y)_x.
\end{eqnarray}

This bracket defines  the $N=2$ supersymmetric Virasoro algebra if we
apply  the Fourier expansion of  $\Phi$ in Eq.(\ref{pois1}).

These supersymmetric equations are possible to obtain from the following Lax representations \cite{LMat,Opop}
\begin{eqnarray} 
&& \alpha = 4, ~~~~~~~ L = -({\cal D}_1{\cal D}_2 + \Phi)^2 ,\hspace{1.6cm}    L_t =4 [ L,L^{3/2}_{\geq 0}] ,               \\  
&& \alpha = -2, ~~~~~ L = \partial^2 + {\cal D}_1 \Phi - {\cal D}_2 \Phi, \hspace{1.1cm} L_t = 4[ L,L^{3/2}_{\geq 0}], \\ 
&& \alpha = 1, ~~~~~~~ L = \partial + \partial^{-1} {\cal D}_1 {\cal D}_2 \Phi, \hspace{1cm}   ~~~ L_t = [L, L^3_{ \geq 1}].
\end{eqnarray}

To this list of  three equations we would like to add the fourth one integrable extension $BN=2$ which has the Lax representation, 
bi-Hamiltonian formulation and possess the superfermionic conserved currents. 

To end this let us send the formula (3) in which $k_i, i=1,2,3,4$ are real coefficients,  to the KdV  Eq.(\ref{kkdv}). Extracting the real 
and imaginary part we obtain the system of two equations on ${(\cal D}_1{\cal D}_2\Phi)_{t}, \Phi_{t,x}$. Solving this system of equations and verifying 
the integrability condition ${(\cal D}_1{\cal D}_2\Phi)_{t,x} = \partial_x \Phi_{t,{\cal D}_1,{\cal D}_2}$ we obtain a system of algebraic equation on the 
coefficients $k_i, i=1,2,3,4$. There are two solutions. 

The first one $k_1=k_4,k_2=-k_3$ is
\begin{equation} 
 \Phi_t = \Phi_{xxx} + 6k_4({\cal D}_1{\cal D}_2\Phi)\Phi_x + 3k_3({\cal D}_1{\cal D}_2 \Phi)^2 - 3k_3\Phi_x^2, \\ 
\end{equation}
and second solution $k_1=-k_4,k_2=k_3$ is
\begin{equation}
\Phi_t = \Phi_{xxx} - 6k_4({\cal D}_1{\cal D}_2\Phi)\Phi_x - 3k_3({\cal D}_1{\cal D}_2 \Phi)^2 + 3k_3\Phi_x^2, \\  
\end{equation} 
The bosonic part of these equations reduces to the KdV equation when $\Phi = \theta_1\theta_2u$ and $k_3=0$.

Therefore without losing on the generality we assume that   $BN=2$ KdV equation has the form 
\begin{equation}\label{comkdv}
\Phi_t = \Phi_{xxx} + 6({\cal D}_1{\cal D}_2 \Phi) \Phi_x.
\end{equation}
Below, we will call such procedure as supercomlexification of the equations.

If we replace  $u$ by  $({\cal D}_1{\cal D}_2 \Phi) + i \Phi_x$  in the Lax operator of KdV equation then we obtain complex operator 
\begin{equation} \label{ckdv1}
 L= \partial_{xx} + ({\cal D}_1{\cal D}_2\Phi) + i \Phi_x
\end{equation}
which  generates the  equation (\ref{comkdv}) but not the conserved currents. We cannot use the supersymmetric version of trace form  Eq.(\ref{trace}) to this operator because 
it does not contain supersymmetric derivatives.   On the other side if we make the same substitution to the conserved currents obtained from the 
Lax operator of KdV equation (\ref{slad}) and make the replacement $\int dx => \int dx d \theta_1 d \theta_2$ in $h_n$ 
\begin{equation} 
 H_n = \int dx h_n => G_n = \int dx d\theta_1 d \theta_2 h_n(u => ({\cal D}_1{\cal D}_2 \Phi) + i \Phi_x), 
\end{equation}
then $G_n = 0$.

The following fourth order  Lax representation, where we do not use the imaginary symbol $i$ generates the supercomplex  $BN=2$ KdV equation. 
\begin{eqnarray} \label{lax}
&&  L = \partial^{4} + 2 \big (\partial ({\cal D}_1{\cal D}_2\Phi)  +
                 ({\cal D}_1{\cal D}_2\Phi)\partial   )\partial  + 
  2 \big (\partial \Phi_x  +  \Phi_{x}\partial  ){\cal D}_1{\cal D}_2,  \\ \label{kadss}
 &&  \hspace{1cm} L_t = 2[L,L^{3/4}_{\geq 0}] ~~~~ => ~~~~~\Phi_t = \Phi_{xxx} + 6({\cal D}_1{\cal D}_2 \Phi) \Phi_x,
 \end{eqnarray} 
The same equation is also generated by the   nonstandard Lax representation 
\begin{eqnarray} \label{lax1}
 && L = \partial + \Phi \partial^{-1}{\cal D}_1{\cal D}_2 + \partial^{-1}(({\cal D}_1{\cal D}_2\Phi) - \Phi {\cal D}_1{\cal D}_2), \\ 
  &&   L_t = [L^{3}_{\geq 1},L] ~~~~ => ~~~~~\Phi_t = \Phi_{xxx} + 6({\cal D}_1{\cal D}_2 \Phi) \Phi_x.
\end{eqnarray}

The  equation  (\ref{kadss}) appeared also for the first time in \cite{Pop2},  where the second Hamiltonian operator was constructed and was interpreted 
as odd version of Virasoro algebra. 

In the components, the equation  (\ref{kadss}) is 
\begin{eqnarray} \label{skdv}
w_{t} &=& w_{xxx} + 6uw_x, ~~~~~ u_t=u_{xxx} + 3(u^2 - w_x^2)_x \\ \nonumber
 \xi_{1,t} &=&   \xi_{1,xxx} + 6\xi_{1,x} u +   6 \xi_{2,x} w_x ,\\ \nonumber  
 \xi_{2,t} &=&   \xi_{2,xxx} - 6\xi_{1,x} w_x + 6\xi_{2,x} u. 
\end{eqnarray}

We see that fermionic sector is invariant under the replacement $ \xi_1 => -\xi_2, \xi_2 => \xi_1$ while the bosonic sector 
is purely even and does not contain the odd functions. This invariance is exactly the $O_2$ invariance. 
Therefore,  this supersymmetric version of the KdV equation   is  $BN=2$  extension.

Introducing  new function $w_x = v$ to  the  bosonic sector of the equation (\ref{skdv}) we obtained
\begin{equation} \label{ckdv} 
 v_t = (v_{xx} +  6 uv )_x, ~~~~
 u_t = (u_{xx} + 3 (u^2 - v^2 ) )_x . 
\end{equation} 
It is exactly the complex KdV equation.  

On the other side the system of equations (\ref{skdv}) is equivalent to the complex version of the Becker and Becker equation Eq.(\ref{becker}).  Indeed if we 
assume that $\Phi= \varrho_1 + i \varrho_2$  in the equation (\ref{becker}) then we obtain 
\begin{eqnarray} \label{cbek}
 \varrho_{1,t} &=& \varrho_{1,xxx} +3\big [{\cal D} (({\cal D}\varrho_1)^2 - ({\cal D}\varrho_2)^2) \big ] \\ \nonumber 
 \varrho_{2,t} &=& \varrho_{2,xxx} +6\big [ {\cal D} (({\cal D}\varrho_1)({\cal D}\varrho_2)) \big ] .
\end{eqnarray}
Next assuming that $ \varrho_1 = \xi_1 + \theta u, \varrho_2= \xi_2 + \theta w, w_x=v $ we see that the previous  equation reduces to the system of equations (\ref{skdv}). 

The Lax representation of the complex KdV equation is given by  the bosonic part of the supercomplexified Lax representation Eq.(\ref{lax}) 

\begin{eqnarray} 
 && L_b= \left ( \begin{array}{cc} \partial^4 + 2( u \partial + \partial u) \partial & 2(\partial v + v\partial) \\
     -2\partial(v\partial + \partial v)\partial & \partial^4 + 2\partial(u\partial + \partial u)                
              \end{array} \right ), \\ \nonumber 
&& ((L^{3/4})_{\geq 0})_b   = \left ( \begin{array}{cc} \partial^3  + 3u\partial & 3v \\ 
                            -3\partial v \partial & \partial^3 + 3\partial u 
                           \end{array} \right ), \\ \nonumber 
~~~~ \\ \nonumber                            
&& \hspace{2cm} L_{b,t} = [ L_b ,((L^{3/4})_{\geq 0})_b],                          
\end{eqnarray}
or by  bosonic part of the supercomplexified Lax representation Eq.(\ref{lax1}) 
\begin{eqnarray} 
 && L_b= \left ( \begin{array}{cc} \partial + \partial^{-1} u  & \partial^{-1} v + v\partial^{-1}, \\ 
     -2 v_x & \partial + u\partial^{-1}  \end{array} \right ),  ~     L_{b,t} = [ L_b , ((L^{3})_{\geq 1})_b].
     \end{eqnarray} 
The symbol $b$ in $L_b$ denotes the bosonic part of the $L$ operator.

There are many differences between the supersymmetric equations (\ref{skdv})  and (\ref{kkdv}). 
The system (\ref{kkdv}) possess an infinite number of conservation laws but the conservation laws of (\ref{skdv}) 
does not reduce to the conservation laws of (\ref{kkdv}). 

Unfortunately if we apply the ``trace form'' to our Lax operator  $ tr ~ (L^{n/4})$ for $n=2,3,5,6$ we did not obtain any conserved currents because then 
 $ tr ~ (L^{n/4})=0$. We confirmed this observation by constructing an arbitrary superbosonic functions  $K_n=K_n(\Phi, ({\cal D}_1\Phi),({\cal D}_2\Phi), ...)  $ 
with  arbitrary constants, where  $n= 1,2,...,11$ denotes the weight. For example 
\begin{eqnarray}
 H_4 &=&  \int ~ dx d\theta_1 d\theta_2 K_4, \\ \nonumber 
 K_4 &=& \lambda_1 \Phi^4 + \lambda_2({\cal D}_1{\cal D}_2\Phi) \Phi^2 + \lambda_3 \Phi_{xx}\Phi + \lambda_4 ({\cal D}_2\Phi)({\cal D}_1\Phi)\Phi ,
\end{eqnarray}
where $\lambda_i, i=1,2,3,4$ are arbitrary constants. 

We  verified that these functions are not constants of motion for our 
supercomplex BN=2 KdV equation.

However, if we expand  $L$ operator as 
\begin{equation} 
 L^{1/8}_1 = {\cal D}_1 + \sum_{k=1}^{\infty} ( \Upsilon_{1,k}  + \varphi_{1,k} {\cal D}_1 + \varphi_{2,k}{\cal D}_2 + 
 \Upsilon_{2,k} {\cal D}_1{\cal D}_2) \partial^{-k},  
\end{equation}
where the super functions $ \Upsilon_{1,k},\Upsilon_{2,k}, \varphi_{1,k},\varphi_{2,k}$ are computed from the assumption that 
$ L =(L^{1/8})^8 $
then  it is possible to obtain the superfermionic conserved currents. 

Indeed,  as we checked    
\begin{equation}
H_{3.5} = tr (L^{7/8})= \frac{1}{4}  \int ~ dx d\theta_1 d\theta_2 \Phi   ({\cal D}_1 \Phi_x) =   -\frac{1}{2} \int ~dx (\xi_1 u_x + \xi_2 w_{xx}),
\end{equation} 
is a conserved quantity. 
The lower index in $H_a$ denotes the dimension of the expression $([\theta_1] = -\frac{1}{2}, [\xi_i] = \frac{3}{2})$. 

Using the $O_2$  transformation it is possible to obtain the superpartners of $L^{1/8}_1$ and $ H_{3.5}$ as  $ L^{1/8}_2 = O_2(L^{1/8}), \hat H_{3.5} =O_2( H_{3.5})$ 
\begin{equation} 
 \hat H_{3.5} =  \frac{1}{4}  \int ~ dx d\theta_1 d\theta_2 \Phi   ({\cal D}_2 \Phi_x) =-\int dx ~ (\xi_1 w_{xx} -\xi_2 u_x).
\end{equation}

It is difficult   to obtain the next  conserved quantity using the trace form, because we need to   compute the $L^{1/8}$ 
up to the terms containing $ \partial^{-6} $,  because then $H = tr (L L^{1/4}L^{1/8})$. 

Therefore we assumed the most general form on the next current and verified that 

\begin{eqnarray}  
 H_{5.5} &=& \frac{1}{2} \int ~dx d \theta_1 d \theta_2~~ \Phi \big ( ({\cal D}_1 \Phi_{xxx})   + 
   4 ({\cal D}_1{\cal D}_2 \Phi )({\cal D}_1\Phi_x) +
  4  \Phi_{xx} ({\cal D}_2 \Phi)  \big ) \\ \nonumber 
 &&=  ~\int ~dx~ \xi_1 \big ( - u_{xx} + 3( w_{x}^2 +  u^2 )  \big )_x  -  \xi_2 \big (- w_{xxx}  - 6uw_x \big )_x,
\end{eqnarray}
is proper conserved quantity.

It is impossible to use the similar trick as in the $BN=1$ supersymmetric KdV equation Eq.(\ref{tryk}) because the transformation  $ 
u =({\cal D}_1{\cal D}_2 \Phi) +  i\Phi_x$ is not invertible. 
Indeed, if we assume that such an inverse exists, then
\begin{equation} \nonumber 
  ({\cal D}_1{\cal D}_2 + i\partial)^2 = 2i \partial({\cal D}_1{\cal D}_2 + i\partial) ~~ =>  ~~ {\cal D}_1{\cal D}_2 + i\partial = 2i\partial,
 \end{equation}
 which is not true. 

Assuming different forms of the Hamiltonian operator of the supercomplexified $BN=2$  KdV equation,  we obtained 
the following bi-Hamiltonian structure 
\begin{eqnarray} 
 && \Phi_t = \frac{1}{2} {\cal D}_1 \partial^{-1} \frac{\delta H_{5.5}  }{\delta \Phi}= \varPi \frac{\delta H_{3.5}}{\delta \Phi}, \\ \nonumber 
 ~~~~&& \\ \nonumber 
  \varPi &=&   2 {\cal D}_1 \partial + 4 \partial^{-1}  [ ({\cal D}_1{\cal D}_2  \Phi){\cal D}_1-   \Phi_x {\cal D}_2 ] + 
 4  [({\cal D}_1{\cal D}_2\Phi) {\cal D}_1 -   \Phi_x {\cal D}_2 ]\partial^{-1}. 
 \end{eqnarray}

We checked that the operator $ \varPi$ defines the proper Hamiltonian operator and satisfies the Jacobi identity 

\begin{equation}
\int dx ~ d \theta_1 d \theta_2 ~~ \alpha ~\varPi^{\star}_{\varPi \beta} ~\gamma ~~ + ~~ cyclic(\alpha,\beta,\gamma) =0, 
\end{equation}

where $\varPi^{\star}_{\varPi \beta } $ is a Gateaux derivative \cite{Blask}
\begin{equation}
 \varPi^{\star}_{\varPi \beta}= \big [\frac{d}{d \epsilon}\varPi(\Phi + \epsilon \varPi \beta) \big]_{\epsilon=0},
\end{equation}
and $\alpha, \beta ,\gamma$ are superfermionic test functions. 
The operator $\varPi$ appeared  first time in \cite{Pop2} and has been connected with the odd version of the Virasoro algebra. 
\begin{equation} \label{pois}
 \{ \Phi(x,\theta_1, \theta_2),\Phi(y,\theta_1^{'}, \theta_2^{'}\} = \varPi \delta(x-y)(\theta_1 - \theta_1^{'})(\theta_2 - \theta_2^{'}).
\end{equation}
This  could be rewritten in the components as
\begin{eqnarray} \label{ody} 
&& \{ w(x),w(y) \} = \{w(x), u(y) \} = \{u(x),u(y) \} =0, \\ \nonumber 
&& \{ w(x),\xi_1(y) \} =-(4w_x\partial^{-1}  + 2\partial^{-1}w_x) ) \delta(x-y), \\ \nonumber 
&& \{ w(x), \xi_2(y)\} = - (2\partial + 4 (\partial^{-1}  u + u\partial^{-1})) \delta(x-y), \\ \nonumber 
&& \{ \xi_1(x),u(y) \} = 2( \partial^2 + 4u - 2\partial u_x) \delta(x-y) , \\ \nonumber 
&& \{ \xi_2(x),u(y) \} = - 4(w_x - \partial^{-1} w_{xx}) \delta(x-y), \\ \nonumber 
&& \{ \xi_1(x), \xi_1(y) \} = -\{ \xi_2(x),\xi_2(y) \} = -4 (\xi_1 \partial^{-1} + \partial^{-1} \xi_1 ) \delta(x-y), \\ \nonumber 
&& \{ \xi_1(x),\xi_2 \} = -4(\xi_{2,x} \partial^{-1} + \partial^{-1}\xi_{2,x} ) \delta(x-y).
\end{eqnarray}

If we compare the right hand  of these   Poisson brackets with the right hand  of Poisson brackets connected with  the $N=2$ supersymmetric 
Virasoro algebra Eq.(\ref{comvir}) we see that they are in opposite statistics. Indeed the supersymmetric 
Poisson brackets (\ref{comvir}) can be symbolically rewritten  as
\begin{equation} 
 \{ B,B\} => {\cal B},~~~ \{ F, B\} => {\cal F} , ~~~~\{ F, F\} => {\cal B}, 
\end{equation}
while the brackets (\ref{ody}) as
\begin{equation} 
 \{ B,B\} => 0 ,~~~ \{ F, B\} => {\cal B} , ~~~~\{ F, F\} => {\cal F},  
\end{equation}
where $B$ denotes $u$ or $w$, $F$ denotes $\xi_1$ or $\xi_2$, and ${\cal B}$ denotes some bosonic operator while ${\cal F}$ denotes some fermionic operator. 
The brackets (\ref{ody}), according to general theory of odd Poisson brackets \cite{sorok},  meets the following properties 
\begin{eqnarray} 
&& \{ A,B+C \} =  \{A,B \} + \{A, C\}, \\
&& \{A,BC \} = \{A,B \}C + (-1)^{(g(A)+1)g(B)}  B \{A, C\} , \\
&&   (-1)^{(g(A) + 1)(g(B)+1)} \{A,\{B,C \} \} + (-1)^{(g(B) + 1)(g(C)+1)} \{B,\{C,A \} \}+ \\ \nonumber 
&& \hspace{2cm} (-1)^{(g(C) + 1)(g(A)+1)} \{C,\{A,B \} \}=0,
 \end{eqnarray} 
where A,B,C are homogeneous elements of the Poisson algebra, $g(A)$ denotes the parity of A. The last equation is a generalized Jacobi identity.

\vspace{0.4cm} 

Instead of using the recursion operator to generate  the conserved currents it is possible to join the superfermionic currents with 
the usual conserved currents of KdV equation using the formula 
\begin{equation} \label{supmod}
 H =\int_0^{1}  ~d\lambda \int dx ~d\theta_1~d\theta_2 \Phi \frac{\delta ({\cal D}_1^{-1}\hat h_{kdv}) }{\delta \Phi}, 
\end{equation}
where $\hat h_{kdv}$ is some conserved currents of KdV equation  in which we make the replacement  
$ u =>  \lambda(({\cal D}_1{\cal D}_2 \Phi) + i \Phi_x)$. If we split the conserved currents $H$   onto the real and imaginary part as
$ H = G + i\hat G$ then it appears  that $\hat G =- O_2(G) $ .

Using the previous formula,  we obtained next conserved current for the supercomplexified KdV equation
\begin{eqnarray}
 H_{7.5} &=&  \int dx ~d \theta_1~d \theta_2~~\Phi \Big ( 3({\cal D}_1 \Phi_{5x}) + 
[  20({\cal D}_1 \Phi_{xx}) ({\cal D}_1{\cal D}_2 \Phi) ]_x     + \\ \nonumber 
 &&[ 20 ({\cal D}_2 \Phi_{xx})\Phi_{x} ]_x + 
 45 ({\cal D}_1\Phi_x) [ ({\cal D}_1{\cal D}_2\Phi)^2  - \Phi_x^2 ] + 10({\cal D}_2\Phi_x)[  
 9 ({\cal D}_1{\cal D}_2\Phi)\Phi_x + 2\Phi_{xxx}] \Big ). 
 \end{eqnarray}

\vspace{0,7cm} 
   
To finish this section let us notice that substitution (3) suggests to replace r.h.s of (3) by complex  chiral  superfield $F$ 
\cite{topa,krzywy} as 
\begin{equation} \label{kiryca}
 u =>  (D F) + (\overline{D} \overline{ F}),
 \end{equation}
where 
\begin{eqnarray} 
 F &=& \xi + \theta_2 u - \frac{1}{2} \theta_1 \theta_2 \xi_x , \hspace{1cm}   
 \overline{F} =  \overline{\xi} +\theta_1 \overline{u} + \frac{1}{2} \theta_1\theta_2 \overline{\xi}_x,  \\
 D &=&  \frac{\partial }{\partial \theta_1} + \frac{1}{2} \theta_2 \partial_x, 
 \hspace{2cm} \overline{D} =\frac{\partial}{\partial \theta_2} + \frac{1}{2} \theta_1 \partial_x , 
\end{eqnarray}
u  is a complex function, $\xi $ is  Grassman valued function. 

Substituting formula (\ref{kiryca}) to the KdV equation we obtain 
\begin{equation} 
 F_t = F_{xxx}  + 6F_x [(DF) + (\overline{D} \overline{F}) ].
\end{equation}
The bosonic part of this equation is 
\begin{equation} 
 u_t = u_{xxx} + 6u_xu + 6\overline{u} u_x - 6 \xi_x\overline{\xi}_x.
\end{equation}
But it is not  the complex KdV equation when $u = a+ib$ where $a,b$ are real functions.

 \section{N=1, BN=1 and BN=2 supersymmetric Sawada-Kotera equation} 
 The Sawada-Kotera equation could be obtained from the  Lax representation 
 \begin{eqnarray} \label{skk}
  L &=& \partial^{3} + u\partial, ~~~~~~~L_t =9 [L, L^{5/3}_{\geq 0}],  \\ 
  &&  u_t=   u_{5x} + 5u_{xxx}u +5u_{xx}u_x + 5u_xu^2  .
\end{eqnarray} 
  The  bi-Hamiltonian formulation  of this equation is 
 \begin{eqnarray} 
  &&u_t = \frac{1}{6} (\partial^{3} + 2(\partial u + u \partial))\frac{\delta G_6}{\delta u},  \\ 
 && \frac{1}{2}  \big (2\partial^{3} + 2u\partial + 2\partial u +  \partial^{-1}(2u_{xx} + u^2) + 
  (2u_{xx} + u^2)\partial^{-1} \big ) u_t = \frac{\delta G_{12}}{\delta u} \\ \nonumber 
  G_6 &= &  \int dx ~(3 u u_{xx} + u^3), \\ \nonumber 
  G_{12} &=&  \frac{1}{18} \int dx ~ u( 9u_{8x} + 96u_{4x}u_{xx}  + 33u_{xxx}^2 + 144u_{xx}^2u + 153u_{xx}u_x^2 - 150u_x^2u^2 + 4u^5). 
    \end{eqnarray}

\subsection{ N=1 , BN=1 susy Sawada-Kotera equation}
 
  The $N=1$  supersymmetric extension of S-K equation is defined by the Lax operator $L = ({\cal D}\partial+  \Phi)^2 $ and its 
  Lax representation \cite{Liu1} 
  \begin{eqnarray} 
  && L_t = [ L,L^{5/3}_{\geq 0} ]~~ => ~~\\ 
 &&   \Phi_t = \frac{1}{9} (\Phi_{5x} + 5\Phi_{xxx}({\cal D}\Phi)+ 5\Phi_{xx}({\cal D} \Phi_x) + 5\Phi_x ({\cal D}\Phi)^2), \\ \nonumber 
 &&  \xi_t = \frac{1}{9} \big ( \xi_{5x}  + 5\xi_{xxx} u + 5\xi_{xx} u_x + 5\xi_x u^2\big ), \\ \nonumber 
  &&  u_t = \frac{1}{9} \big ( u_{5x} +  5u_{xxx} u  + 5u_{xx}u_x + 5u_x u^2 - 5\xi_{xxx}\xi_x \big ) . 
  \end{eqnarray}

The odd bi-Hamiltonian representation  for  supersymmetric $N=1$ extension of the Sawada-Kotera equation has been given in \cite{Pop4} 
\begin{eqnarray} 
 \Phi_t &=&  ({\cal D} \partial^{2} + 2\partial \Phi + 2\Phi \partial + {\cal D}\Phi {\cal D})  \partial^{-1} 
  ( {\cal D} \partial^{2} + 2\partial \Phi + 2\Phi \partial + {\cal D} \Phi {\cal D}) \frac{\delta H_4}{\delta \Phi},  \\ 
  && ~~~~~~ \big ( \partial^{2} + ({\cal D}\Phi) - \partial^{-1}{\cal D}\Phi_x  + \Phi_x\partial^{-1}{\cal D}\big ) \Phi_t = 
  \frac{\delta H_{10}}{\delta u},  \\ \nonumber 
  H_4 & =& \frac{1}{18} \int dx~d\theta ~\Phi \Phi_x , \\ \nonumber 
  H_{10} &=& \frac{1}{54} \int dx ~ d\theta~ \Phi  \Big [-3\Phi_{7x} + 12 \Phi_{5x} ({\cal D}\Phi) + 28\Phi_{4x} ({\cal D}\Phi_x) + \\ \nonumber 
&&   3\Phi_{xxx}\big [32({\cal D}\Phi_{xx}) + 15({\cal D}\Phi)^2\big ] +
   \Phi_{xx}\big [8({\cal D}\Phi_{xxx})+ 30({\cal D}\Phi_x)({\cal D}\Phi)\big ] + \\ \nonumber 
  &&  \Phi_x\big [4({\cal D}\Phi_{4x}) + 30({\cal D} \Phi_{xx})({\cal D}\Phi )\big ] + 
  15({\cal D}\Phi_x)^2 + 8({\cal D}\Phi)^3 \Big ]. 
\end{eqnarray}

 The $BN=1$ supersymmetrical S-K equation is defined by the Lax operator  $ L = \partial^{3} + ({\cal D}\Phi)\partial $ and its Lax representation as
   \begin{eqnarray} 
&&  L_t = [ L,L^{5/3}_{\geq 0} ] => 
   \Phi_t = \frac{1}{9} (\Phi_{5x} + 5\Phi_{xxx}({\cal D}\Phi)+ 5\Phi_{x}({\cal D} \Phi_{xx} + 5\Phi_x ({\cal D}\Phi)^2), \\ \nonumber 
&&   \xi_t = \frac{1}{9} \big ( \xi_{5x}  + 5\xi_{xxx} u + 5\xi_{x} u_{xx} + 5\xi_x u^2\big ), \\ \nonumber 
&&    u_t = \frac{1}{9} \big ( u_{5x} +  5u_{xxx} u  + 5u_{xx}u_x + 5u_x u^2  \big ).  
  \end{eqnarray}

The bi-Hamiltonian formulation for the $BN=1$ extension is easy to obtain using the same trick as in the case of the $BN=1$ extension of KdV equation,  
see Eq. (\ref{tryk}).

\subsection{ BN=2 Supercomplex  Sawada-Kotera equation}

The following Lax operator 
\begin{equation} \label{laxsw} 
 L = \partial^{3} + (k_1 ({\cal D}_1 {\cal D}_2 \Phi) + k_2 \Phi_x)  \partial + 
 (-k_2({\cal D}_1{\cal D}_2 \Phi)  + k_1\Phi_x ) {\cal D}_1 {\cal D}_2, 
\end{equation}
where $k_1, k_2$ are  arbitrary constants, generates  the $BN=2$ supersymmetrical  Sawada-Kotera equation
\begin{eqnarray} \label{bn2} 
L_t &=& 9  [L,L^{5/3}_{\geq 0} ], \\  \label{laksw}
\Phi_t &=& \frac{1}{3} \big [ 3\Phi_{5x} 
+ 15 ({\cal D}_1 {\cal D}_2 \Phi_{xx}) (k_1\Phi_x - k_2 ({\cal D}_1{\cal D}_2\Phi) )  - 10k_1k_2({\cal D}_1{\cal D}_2\Phi)^3 +\\ \nonumber 
&&  15 ({\cal D}_1 {\cal D}_2 \Phi)^2\Phi_x(k_1^2 - k_2^2) + 15k_1({\cal D}_1{\cal D}_2\Phi)( \Phi_{xxx}  + 2k_2\Phi_x^2)  +\\ \nonumber 
&& 15k_2 \Phi_{xxx} \Phi_x + 5\Phi_x^3(k_2^2 - k_1^2)   \big ].
\end{eqnarray}
This equation is also possible to obtain  after  modification of the supercomplexification method as 
\begin{equation} \label{kotek} 
 u => k_1 ({\cal D}_1{\cal D}_2 \Phi) + k_2 \Phi_x + i(-k_2 ({\cal D}_1{\cal D}_2 \Phi) + k_1\Phi_x) 
\end{equation}
and substituting it to the Sawada- Kotera  equation or to the Lax representation Eq.(\ref{skk}).  
However then  our  Lax operator is a complex operator which does not generate the conserved currents.

Introducing a new function $w_x = v$,  it appears that it is always possible to find the linear transformation of  $v,u$   which  
changes the bosonic sector of the equation (\ref{bn2}) to the complex version of the Sawada-Kotera equation for any arbitrary values of $k_1,k_2$,
\begin{eqnarray} 
&&  v_t = \frac{1}{3} \big [ 3v_{4x} + 15v_{xx}u - 5v^3 + 15vu_{xx} + 15vu^2 \big ]_x, \\ \nonumber 
&&  u_t = \frac{1}{3} \big [ 3u_{4x} + 15u_{xx}u + 5u^3 - 15v_{xx}v - 15v^2u \big ]_x.
\end{eqnarray}

In order to study the conservation laws and Hamiltonian structure of the $BN=2$ supersymmetric Sawada-Kotera equation, we consider the special case $ k_1=1,k_2=0$ 
for which we obtained
\begin{eqnarray} 
&&   L = \partial^{3} + ({\cal D}_1 {\cal D}_2 \Phi) \partial + \Phi_x {\cal D}_1 {\cal D}_2, \\
&& \Phi_t = \frac{1}{3} \big [ 3\Phi_{5x} + 15 ({\cal D}_1 {\cal D}_2 \Phi_{xx}) \Phi_x +   
  15 ({\cal D}_1 {\cal D}_2 \Phi)^2\Phi_x + \\ \nonumber 
&& \hspace{2cm}   15({\cal D}_1{\cal D}_2\Phi)\Phi_{xxx}  -5 \Phi_x^3 \big ].\label{bn3sk}
\end{eqnarray}

In the components,  the equation Eq.(\ref{bn3sk}) is 
\begin{eqnarray}\label{n2saw}
\Phi &=& w + \theta_1 \xi_1 + \theta_2 \xi_2 + \theta_1\theta_2 u , \\ \nonumber 
w_t &=& \frac{1}{3} \big [  3w_{5x}+15w_{xxx}u -5w_x^2  + 15w_{x}u_{xx} + 15w_xu^2 \big ] ,  \\ \nonumber 
~~~\\ \nonumber 
u_t &=& \frac{1}{3} \big [ 3u_{4x} +  15u_{xx}u + 5u^3 - 15 w_{xxx}w_x -15w_x^2u  \big ]_x , \\ \nonumber 
~~\\ \nonumber 
\xi_{1,t} &=&\big [ \xi_{1,5x}+ 5\xi_{1,xxx}u  +5 \xi_{1,x}(u_{xx} + u^2  - w_x^2 ) + 
  \xi_{12,xxx}w_{x}  +  5\xi_{2,x}( w_{xxx} + 2w_xu) \big ], \\ \nonumber
~~\\ \nonumber 
 \xi_{2,t}& =& \big [ \xi_{2,5x}+ 5\xi_{2,xxx}u + 5\xi_{2,x}(u_{xx} + u^2 - w_x^2) -
 5\xi_{1,xxx}w_{x} - 5\xi_{1,x}(w_{xxx} + 2w_x u^2) \big ]. 
\end{eqnarray}

The bosonic sector of the Eq. (\ref{n2saw})  does not interact with the fermionic variables. Thus we have  
the $BN=2$ extension of the Sawada-Kotera equation.

In order to find the conserved current,  we use the formula  Eq.(\ref{supmod}) 
\begin{equation} 
 H =\int_0^{1}  ~d\lambda \int dx ~d\theta_1~d\theta_2 \Phi \frac{\delta ({\cal D}_1^{-1}\hat h_{sk}) }{\delta \Phi}, 
\end{equation}
where $\hat h_{sk}$ is some conserved currents of the Sawada-Kotera  equation  in which we make a replacement  
$ u =>  \lambda(({\cal D}_1{\cal D}_2 \Phi) + i \Phi_x)$ . 

Due to this formula  we obtained the following conserved charges
\begin{eqnarray} 
&& H_{5.5}  =  \int ~dx d\theta_1 d\theta_2 \Phi \Big [ 3 ({\cal D}_1\Phi_{xxx}) +
 2({\cal D}_2 \Phi_x) \Phi_x   + 2 ({\cal D}_1\Phi_x) ({\cal D}_1{\cal D}_2 \Phi)  \Big ], \\ \nonumber
&& H_{7.5} = \int ~dx d\theta_1 d\theta_2 \Phi \Big [  ({\cal D}_1\Phi_{5x}) +  2 [ ({\cal D}_2\Phi_{xx})\Phi_x ]_x + 
2({\cal D}_2\Phi) [({\cal D}_1 {\cal D}_2 \Phi )  \Phi_x + \Phi_{xxx} ] + \\ 
&& \hspace{1.5cm} 2 [ ({\cal D}_1\Phi_{xx})({\cal D}_1{\cal D}_2 \Phi)]_x  + ({\cal D}_1\Phi_x) [ 2({\cal D}_1{\cal D}_2 \Phi_{xx}) +
 ({\cal D}_1{\cal D}_2\Phi)^2 -  \Phi_{x}^2 ] \Big ]   \\ 
&&  H_{11.5} =  \int ~dx d\theta_1 d\theta_2 \Phi( 9({\cal D}_1\Phi_{9x})  + 74 ~~terms) . 
\end{eqnarray}

The  system of equation  (\ref{laksw})  could be rewritten as the bi-Hamiltonian system 
\begin{eqnarray} \nonumber
 &&\Phi_t = \frac{1}{6} \Big [  {\cal D}_1 \partial + 
 2 \partial^{-1}[({\cal D}_1{\cal D}_2 \Phi ) {\cal D}_1  - \Phi_x {\cal D}_2] + \\ 
&& \hspace{3cm}  2 [({\cal D}_1{\cal D}_2 \Phi ){\cal D}_1  - \Phi_x {\cal D}_2] \partial^{-1} \Big ]
\frac{\delta H_{5.5}}{\delta \Phi} , \\ 
 && \hspace{3cm}  {\cal K}\Phi_t = \frac{\delta H_{11.5}}{\delta \Phi}, \\ \nonumber 
  {\cal K} = && 18 \Big ({\cal D}_1\partial^4+ 2 \big [ ({\cal D}_2\Phi_x) +  ({\cal D}_1{\cal D}_2\Phi){\cal D}_1 
 + \Phi_x {\cal D}_2 \big ] \partial^2  + \\ \nonumber 
 && \big [({\cal D}_1{\cal D}_2\Phi_x){\cal D}_1+ \Phi_{xx} {\cal D}_2  + ({\cal D}_2\Phi_{xx}) + 2({\cal D}_1\Phi_x) {\cal D}_1{\cal D}_2 \big ] \partial + \\ \nonumber 
 && 2 \big [({\cal D}_2\Phi_{xx}) + ({\cal D}_2\Phi_x) ({\cal D}_1{\cal D}_2\Phi) - ({\cal D}_1\Phi_x)\Phi_x \big ]
+({\cal D}_1\Phi_{xx}){\cal D}_1{\cal D}_2  +  \\ \nonumber 
 &&     \big [({\cal D}_1{\cal D}_2\Phi)^2 -\Phi_x^2 +2 ({\cal D}_1{\cal D}_2\Phi_{xx}) \big ]{\cal D}_1    +      
 2 \big [({\cal D}_1{\cal D}_2\Phi)\Phi_x + \Phi_{xxx} \big ] {\cal D}_2 +  \\ \nonumber 
&& \hspace{1.5cm} {\cal D}_1{\cal D}_2 \partial^{-1} \big [ ({\cal D}_2\Phi_x)\Phi_x +  ({\cal D}_1\Phi_{xxx}) +({\cal D}_1\Phi_x)\Phi_x  \big ] +\\ \nonumber 
&& \hspace{1.5cm} \big [ ({\cal D}_2\Phi_x)\Phi_x +  ({\cal D}_1\Phi_{xxx}) +({\cal D}_1\Phi_x)\Phi_x  \big ]{\cal D}_1{\cal D}_2 \partial^{-1}
\Big ). 
\end{eqnarray}
The operator ${\cal K}$ defines a proper symplectic operator for the $BN=2$ supersymmetric Sawada-Kotera  equation and  satisfies the condition  \cite{Blask}

\begin{equation}  \label{jacobi} 
 \int dx~ d \theta_1~ d\theta_2 ~\big [ \alpha {\cal K}^{\star}_{\beta} \gamma +  \beta {\cal K}^{\star}_{\gamma} \alpha +  
 \gamma  {\cal K}^{\star}_{\alpha} \beta \big ]= 0.
\end{equation}
where $\alpha, \beta,\gamma$ are the test superfunctions  and ${\cal K}^{\star}_{W}$ is a Gateaux derivative defined as 
\begin{equation} \label{gato} 
 {\cal K}^{\star}_{W} = \frac{d}{d \epsilon} K(\Phi + \epsilon W)|_{\epsilon =0}.
\end{equation}
As we checked the equation  (\ref{jacobi}) is satisfied for superfermionic test functions and also for the superbosonic test functions. 
For the superfermionic test functions we should assume that in the formula Eq. (\ref{gato}),  $\epsilon$ is an anticommuting variable, 
$W$ is a superfermionic function  because $\Phi$ is a superbosonic function. 

\noindent  To finish this section  let us mention  that all our formulas presented here possess the $O_2$ superpartners. 
 
 \section{BN=1,BN=2 supersymmetric Kaup-Kupershmidt equation} 
The Kaup-Kupershmidt (K-K) equation is derived from the Lax operator 
\begin{eqnarray} \label{kaupik}
 && L = \partial^3 + \partial u + u\partial, ~~~~ L_t=9 [L,L^{5/3}_{\geq 0}], \\  
 && u_t = u_{5x} + 10u_{xxx}u+25u_{xx}u_x + 20u_x u^2 =(\partial_{xxx} + \partial u + u\partial)  \frac{\delta H_6}{\delta u}, \\ 
 && \Big [ 18 \partial_{xxx} + 90(\partial u + u\partial) + \partial^{-1}[ 144 u^2 + 36 u_{xx}] + \\ \nonumber 
 && \hspace{2cm} [144 u^2 + 36u_{xx}]\partial^{-1} \Big ] u_t = \frac{\delta H_{12}}{\delta u}, \\ 
 && H_6 = \frac{1}{6} \int dx ( 3u_{xx}u + 8u^3),  \\ \nonumber 
 && H_{12} =\int dx (9u_{8x} u - 180u_{xxx}^2 u + 222u_{xx}^3 + 1224u_{xx}^2u^2 - \\  
 && \hspace{3cm} 186u_x^4 -3360u_x^2u^3 + 256u^6).
\end{eqnarray}

 The  $N=1$ supersymmetric extension of the K-K  equation does not exist. 
 It follows from the observation that,  if we assume the most general form on the supersymmetric extension of K-K 
 as the polynomial in $\Phi, ({\cal D}\Phi)$ and its derivatives, 
 which reduces in the bosonic limit to the Kaup-Kupershmidt equation,  then  it is possible to construct 
 only one conserved current. It is not enough for such a  system to  be  integrable . 
 
 However it is possible to obtain the $BN=1$ supersymmetric extension of K-K equation by simply substituting $u=({\cal D}\Phi)$ to 
 the Eq.(\ref{kaupik})
\begin{equation}
 \Phi_t= \Phi_{5x} + 10\Phi_{xxx} ({\cal D}\Phi) + 15\Phi_{xx} ({\cal D}\Phi_x) + 10 \Phi_x({\cal D}\Phi_{xx}) + 20 \Phi_x ({\cal D}\Phi)^2.
\end{equation}

In order to construct the $BN=2$ supersymmetric extension of the Kaup-Kupershmidt equation let us consider the most general supercomplexified 
ansatz 
\begin{equation} 
 u ~~ => ~~  k_1({\cal D}_1{\cal D}_2\Phi) + k_2 \Phi_x + 
 i(k_3({\cal D}_1{\cal D}_2 \Phi) + k_4\Phi_x) , 
\end{equation}
where $k_1,k_2,k_3,k_4$ are arbitrary constants,  and substitute it to the Kaup-Kupershmidt equation.
As a result,  we obtained  $k_3=k_2,k_4=-k_1$  and 
 \begin{eqnarray} 
 &&  \Phi_t =  \Phi_{5x} + 10({\cal D}_1{\cal D}_2\Phi_{xx})(k_1\Phi_x - k_2({\cal D}_1{\cal D}_2\Phi)) + 10k_2\Phi_{xxx}\Phi_x + \\ \nonumber 
 && 15 ({\cal D}_1{\cal D}_2\Phi _x))(k_1 \Phi_{xx} - \frac{1}{2}k_2({\cal D}_1{\cal D}_2\Phi_x))  + \frac{15}{2}k_2 \Phi_{xx}^2 + 
 \frac{20}{3}(k_2^2 - k_1^2)  \Phi_x^3 +\\ \nonumber 
&&   10({\cal D}_1{\cal D}_2\Phi) [ 2(k_1^2 - k_2^2)({\cal D}_1{\cal D}_2\Phi)\Phi_x   -\frac{4}{3}k_1k_2 ({\cal D}_1{\cal D}_2 \Phi)^2  
+ k_1\Phi_{xxx} + 4k_1k_2\Phi_x^2 ]. 
 \end{eqnarray}

 It is possible to transform the bosonic part of $\Phi_t$ to the complex Kaup-Kupershmidt equation after the identification $w_x=v$ and after making the linear 
 transformation of the function $v,u$ for arbitrary  values of $k_1,k_2$ 
 \begin{eqnarray}
 &&  v => \frac{k_2}{k_1^2 + k_2^2} u + \frac{k_1}{k_1^2 + k_2^2}v, ~~~~~~~ u => \frac{k_1}{k_1^2 + k_2^2} u - \frac{k_2}{k_1^2 + k_2^2}v, \\ \nonumber 
 &&  u_t = u_{5x} + 10u_{xxx}u + 25u_{xx}u_x + 20u_xu^2 - 10v_{xxx}v - 5v_x(5v_{xx} + 8vu) - 20v^2u_x,\\ \nonumber 
 &&  v_t = v_{5x} + 10 v_{xxx}v + 25v_{xx} u_x  + 25v_xu_{xx}  +10v(u_{xxx} + 4u_xu)  + 5v_x( 5u_{xx} + 4u^2 - 4v^2).
 \end{eqnarray}
Without losing on the generality,  we assume  that $k_2=0,k_1=1$  and hence we consider following equation 
 \begin{eqnarray} \label{kaup1}
 &&  \Phi_t =  \Phi_{5x} + 10({\cal D}_1{\cal D}_2\Phi_{xx})\Phi_x + 
 15 ({\cal D}_1{\cal D}_2\Phi _x)\Phi_{xx} + \\ \nonumber  
&& \hspace{1cm}  10({\cal D}_1{\cal D}_2\Phi)(2 ({\cal D}_1{\cal D}_2\Phi)\Phi_x  + \Phi_{xxx}) - \frac{20}{3} \Phi_{x}^3 .
 \end{eqnarray}
In the components the Eq. (\ref{kaup1}) is
 \begin{eqnarray}
\Phi &=& w + \theta_1 \xi_1 + \theta_2 \xi_2 + \theta_1\theta_2 u , \\ \nonumber 
w_t &=& \frac{1}{3} \big [ 3w_{5x}+30w_{xxx}u + 45w_{xx}u_x - 20w_x^3  +30w_x(u_{xx} - 2u^2) \big ],  \\ \nonumber 
~~~\\ \nonumber 
u_t &=& \frac{1}{6} \big [ 6u_{4x} +  60u_{xx}u + 45u_x^2 + 40u^3 -  60 w_{xx}w_x  -45w_{xx}^2 - 120w_x^2u   \big ]_x , \\ \nonumber 
~~\\ \nonumber 
\xi_{1,t} &=&\big [ \xi_{1,5x}+ 10\xi_{1,xxx}u  +1 5 \xi_{2,xxx}u_x + 10(\xi_{2,x}w_x)_{xx} +  \\ \nonumber
&& \hspace{1.5cm}  10\xi_{1,x}(u_{xx} + 2u^2 - 2w_{xx}^2)  +  40\xi_{2,x}w_{x}u  \big ],\\ \nonumber
~~\\ \nonumber 
\xi_{2,t} &=& \big [ \xi_{2,5x} -105\xi_{1,xxx}w_x - 15\xi_{1,xx}w_{xx} - 10 \xi_{1,x}w_{xxx} -40\xi_{1,x}w_xu + \\ \nonumber
&& \hspace{1cm} 10 \xi_{2,xxx}u +15\xi_{2,xx}u_{x} + 10\xi_{2,x}(u_{xx} + 2u^2 -2w_x^2) \big ].
\end{eqnarray}

In order to find the conserved current for $BN=2$ supersymmetric Kaup-Kupershmidt equation, we apply the same method as  used in the supersymmetric 
$BN=2$ Sawada-Kotera equation. 
Therefore, we apply  the formula  Eq.(\ref{supmod}) in which now 
\begin{equation} 
 H =\int_0^{1}  ~d\lambda \int dx ~d\theta_1~d\theta_2 \Phi \frac{\delta ({\cal D}_1^{-1}\hat h_{kk}) }{\delta \Phi}, 
\end{equation}
where $\hat h_{sk}$ is some conserved current of the Kaup-Kupershmidt   equation  in which we make the replacement  
$ u =>  \lambda(({\cal D}_1{\cal D}_2 \Phi) - i \Phi_x)$.
As a result we obtained the following conserved currents 
\begin{eqnarray} 
&& H_{5.5} =\int~dx d\theta_1 d\theta_2\Phi \Big [ 3 ({\cal D}_1\Phi_{xxx}) +
 16({\cal D}_2 \Phi_x)\Phi_x + 16 ({\cal D}_1\Phi_x) ({\cal D}_1{\cal D}_2 \Phi)], \\ \nonumber 
&& H_{7.5} = \int ~dx d\theta_1 d\theta_2 \Phi \Big [({\cal D}_1{\cal D}_2\Phi_{5x}) +
8(({\cal D}_2\Phi_{xx}) \Phi_x)_x  +\\
&& 8(({\cal D}_1\Phi_{xx})({\cal D}_1{\cal D}_2 \Phi)_x + 
 8({\cal D}_2\Phi_x)(4({\cal D}_1{\cal D}_2\Phi)\Phi_x + \Phi_{xxx}) + \\ \nonumber 
&& \hspace{2cm} 8({\cal D}_1\Phi_x) (({\cal D}_1{\cal D}_2\Phi_{xx}) + 2({\cal D}_1{\cal D}_1\Phi)^2 - 2\Phi_x^2)  \Big ] \\ 
&& H_{11.5} = \int ~dx d\theta_1 d\theta_2  \Phi(9({\cal D}_1\Phi_{9x})  + 74 ~~terms). 
\end{eqnarray}

Now the bi-Hamiltonian formulation is 
\begin{eqnarray} \nonumber 
 && \Phi_t = \\ 
 && \frac{1}{6} \Big ( {\cal D}_1\partial + \partial^{-1} [ ({\cal D}_1{\cal D}_2\Phi) {\cal D}_1 - \Phi_x{\cal D}_2] + 
 [({\cal D}_1{\cal D}_2\Phi) {\cal D}_1 - \Phi_x{\cal D}_2 ]\partial^{-1} \Big )\frac{\delta H_{5.5}}{\delta \Phi} \\ 
 &&  \hspace{3cm} {\cal K} \Phi_t  =   \frac{\delta H_{11.5}}{\delta \Phi} \\ \nonumber 
 && {\cal K} = 18 \Big [ {\cal D}_1\partial^4 + 10 \big [{\cal D}_2\Phi_{x} + ({\cal D}_1{\cal D}_2 \Phi){\cal D}_1 \big ]\partial^2  + 5 \big [({\cal D}_1{\cal D}_2 \Phi_x){\cal D}_1 
 + {\cal D}_2 \Phi_{xx} + \\ \nonumber 
 &&  2({\cal D}_1\Phi_x) {\cal D}_1{\cal D}_2 \big ] \partial + 4 \big [ ({\cal D}_2 \Phi_{xxx}) + 8({\cal D}_2\Phi_x) ({\cal D}_1{\cal D}_2 \Phi ) - 
 8({\cal D}_1\Phi_x) \Phi_x \big ] + \\ \nonumber 
&&  4 \big [  ({\cal D}_1{\cal D}_2 \Phi_{xx}) + 4({\cal D}_1{\cal D}_2\Phi)^2 -  4 \Phi_x^2  \big ] {\cal D}_1 +
4\big [8({\cal D}_1{\cal D}_2\Phi) \Phi_x + \Phi_{xxx} \big ] {\cal D}_2 + \\ \nonumber 
&& 2 \big [ 8({\cal D}_2\Phi_x)\Phi_x + ({\cal D}_1\Phi_{xxx}) +  8({\cal D}_1\Phi_x) ({\cal D}_1{\cal D}_2\Phi) \big ]\partial^{-1} {\cal D}_1{\cal D}_2 +\\ \nonumber 
&&  2 {\cal D}_1{\cal D}_2 \partial^{-1} \big [ 8({\cal D}_2\Phi_x)\Phi_x + ({\cal D}_1\Phi_{xxx}) +  8({\cal D}_1\Phi_x) ({\cal D}_1{\cal D}_2\Phi) \big ] \Big ]
\end{eqnarray}

The operator ${\cal K}$ defines a proper symplectic operator for the $BN=2$ supersymmetric Kaup-Kupershmidt equation and  satisfies the condition  \cite{Blask}
\begin{equation} 
 \int dx~ d\theta_1~ d \theta_2 ~\big [ \alpha {\cal K}^{\star}_{\beta} \gamma +  \beta {\cal K}^{\star}_{\gamma} \alpha +  
 \gamma  {\cal K}^{\star}_{\alpha} \beta \big ]= 0.
\end{equation}

To finish this section,  let us notice that all our formulas possess the $O_2$ superpartners. 

\section{Conclusion} 
 In this paper, the method of the  $BN=2$  supercomplexification   has been applied to the supersymmetrization of known soliton equations. 
 In that manner,  we obtained new supersymmetric KdV equation  with its odd bi-Hamiltonian and Lax representation. Also,  the $BN=2$ supercomplexification of 
 the Sawada-Kotera with its Lax representation and Kaup-Kupershmidt equations have been discussed. Unfortunately,  we have been not able to find Lax 
 representation for the $BN=2$ Kaup-Kupershmidt equation.
 The unexpected feature of the supercompexification is appearance of the odd Hamiltonians 
 operators and superfermionic conserved currents. The $O_2$ invariance of the  conserved currents and Hamiltonian operators has a  special meaning here. 
It is similar  to the invariance of the  conserved currents in the complex soliton system. For example,  plugging the function $u  => u+ iv$ to  
some conserved current
 $H=H(u,u_x,\dots)$ we obtain  $H => H_r + iH_i$ where  $H_r$ and $H_i$ are conserved too. In the $N=2$ supercomplex version  if $H$ 
 is conserved then $O_2(H)$ is also conserved. 
 The supersymmetric Lax operator,  which generates the $BN=2$ supercomplex KdV equation, generates also the superfermionic conserved currents. The bosonic part of this
 Lax operator generates the complex KdV equation.  However, we do not know how it is possible to obtain  the conserved currents of complex KdV equation using 
 this  operator.
 On the other hand, it seems that the supercomplexification is a general method and could be applied to  wide classes of 
 integrable equations. 
 
\section{Acknowledgements} 

I would like to thank the anonymous referees for the constructive remarks.

\end{document}